\documentclass[pre,reprint,floatfix,superscriptaddress,showpacs,amsmath,amssymb]{revtex4-1}
\usepackage{graphicx}
\usepackage{bm}

\usepackage{color,ulem}

\usepackage{amsmath}	
\usepackage{mathrsfs}%
\begin{document} 

\title{Coarsening dynamics driven by vortex-antivortex annihilation
in ferromagnetic Bose-Einstein condensates} 
\author{Kazue Kudo} 
\affiliation{Department of Computer Science, 
Ochanomizu University, 2-1-1 Ohtsuka, Bunkyo-ku, Tokyo 112-8610, Japan} 
\author{Yuki Kawaguchi} 
\affiliation{Department of Applied Physics, 
University of Tokyo, 7-3-1 Hongo, Bunkyo-ku, Tokyo 113-8656, Japan} 
 
\date{\today} 
\begin{abstract} 
 In ferromagnetic Bose-Einstein condensates (BECs), 
 the quadratic Zeeman effect controls magnetic anisotropy, which affects
 magnetic domain pattern formation.
 While the longitudinal magnetization is dominant (similar to the Ising
 model) for a negative quadratic Zeeman energy, the transverse
 magnetization is dominant (similar to the $XY$ model) for a positive
 one. When the quadratic Zeeman energy is positive, the coarsening
 dynamics is driven by vortex-antivortex annihilation in the same way as
 the $XY$ model. However, due to superfluid flow of atoms, there exist
 several 
 combinations
 of vortex-antivortex pairs in ferromagnetic BECs, which
 makes the coarsening dynamics more complicated than that of the $XY$
 model.
 We propose a revised domain growth law, which is based on the growth
 law of the
 two-dimensional $XY$ model, for a two-dimensional ferromagnetic BEC with
 a positive quadratic Zeeman energy. 
\end{abstract} 

\maketitle 
 
\section{Introduction}
\label{sec:intro}

Domain growth and coarsening dynamics have been studied in a wide 
variety of systems
\cite{Bray,Lifshitz,Ohta,Huse, Siggia,Furukawa,
Bray90,Pargellis,Yurke,Puri,Rutenberg}.  
When a system is quenched from a disordered phase to an
ordered phase, the long-range order does not arise immediately.
At first, locally-ordered small domains arise, which then grow with time
to make global order through domain coarsening.   
Although there are different mechanisms to cause domain growth, 
in most cases,  
domain size $l$ grows with time $t$ as $l(t)\sim t^\nu$, where $\nu$ is
an scaling exponent. 
For example, $\nu=1/3$ in the two-dimensional (2D) conserved systems
described by the Ising model (e.g., binary alloys and ferromagnets
with uniaxial anisotropy)~\cite{Bray,Lifshitz,Ohta,Huse}.  
If fluid flow contributes to domain growth, which is the case of binary
fluids, the exponent changes depending on advection and viscosity. 
When the advective transport is negligible, diffusion dominates the
coarsening dynamics. In that case, $\nu=1/3$~\cite{Siggia}, which is
the same as in the absence of flow. 
However, if the advective transport with little viscosity dominates
over diffusion,  
the inertia of fluid becomes important in the coarsening dynamics.
In this case, domains grow faster than the diffusive case,
and the exponent is $\nu=2/3$~\cite{Furukawa}.
When the system is described by vector fields (i.e., complex order
parameters or multi-component order parameters), 
the dominant mechanism to cause domain growth is completely different
from those for the Ising model and binary fluids.
For example, the coarsening dynamics for 2D vector fields is driven by
the annihilation of vortex-antivortex pairs.  
The domain size, which is actually the characteristic length of the
spatial structure of the field, grows as $l(t)\sim t^{1/2}$ for
non-conserved $n$-component vector fields in $d$-dimensional space,
except for $d=n=2$, namely, the 2D $XY$ model.
The domain growth law for the 2D $XY$ model includes a logarithmic
correction:  
$l(t)\sim (t/\ln t)^{1/2}$~\cite{Bray90,Pargellis,Yurke,Puri,Rutenberg}.

Magnetic domain patterns and their coarsening dynamics are observed also
in ferromagnetic Bose-Einstein condensates (BECs).
Recent development in imaging techniques to observe magnetization
profiles in ferromagnetic BECs has enabled us to investigate the
real-time dynamics of magnetization, such as spin texture formation,
spin-domain coarsening, 
and nucleation of spin 
vortices~\cite{Sadler2006,berkeley08,Vengalattore2010,De2014}.  
Those experiments have also motivated theoretical studies about 
configurations of Skyrmions and spin
textures~\cite{Barnett2009,Cherng2011},
magnetic domain formation~\cite{Kudo11,Kudo13}, and
spin turbulence~\cite{Fujimoto12,Tsubota13,Fujimoto13}.
Magnetic anisotropy of a ferromagnetic BEC depends on the quadratic
Zeeman energy, which can be controlled by external fields.
When the quadratic Zeeman energy is negative, longitudinal magnetization
is dominant, and thus the system is similar to the Ising model.
In 2D ferromagnetic BECs with a negative quadratic Zeeman
energy or binary BECs,
domain size grows as $l(t)\sim t^{2/3}$~\cite{Kudo13,Hofmann}, 
which has the same exponent $\nu=2/3$ as that for binary fluids in
the inertial hydrodynamic regime.   
However, $l(t)\sim t^{1/3}$ in the absence of superfluid
flow~\cite{Kudo13}. The difference in the exponents suggests that the
superfluid flow has a strong influence on the coarsening dynamics.

In this paper, we investigate the coarsening dynamics in 2D 
spin-$1$ ferromagnetic
BECs with a positive quadratic Zeeman energy. When the quadratic Zeeman
energy is positive, transverse magnetization is dominant, and the
coarsening dynamics is caused by vortex-antivortex annihilation.
The situation is similar to the 2D $XY$ model, however a crucial
difference arises in the classification of vortices.
Vortices in ferromagnetic BECs are classified by the winding number of
spin current (direction of magnetization) and 
mass current (vorticity of superfluid flow). 
Thus, there are several combinations
of vortex-antivortex pairs which cause
pair annihilation in ferromagnetic BECs. 
By contrast, in the $XY$ model, there is only one 
combination of vortex-antivortex pairs.
When several combinations of annihilation pairs exist,
the coarsening dynamics 
is expected to be more complicated than that of the $XY$
model. In other words, superfluid flow has indirect effects on the coarsening
dynamics through different 
combinations of vortex-antivortex annihilation. 
We will demonstrate the coarsening dynamics in ferromagnetic BECs by
numerical simulations, and propose a revised domain growth law, based on
the growth law for the $XY$ model. 

The rest of the paper is organized as follows. 
The decay of the vortex density, which is caused by vortex-antivortex
annihilation, in ferromagnetic BECs is discussed in
Sec.~\ref{sec:law}. Numerical simulations illustrated in
Sec.~\ref{sec:simu} clearly show that superfluid flow affects the
coarsening dynamics and that a revised growth law is needed for the case
where there are several 
combinations of vortex-antivortex pairs. 
The revised law is proposed in Sec.~\ref{sec:rev}.
Discussions and conclusions are given in Sec.~\ref{sec:disc}.

\section{Domain growth law}
\label{sec:law}

\subsection{Interaction of vortices}
\label{sec:vortex}

We consider a spin-1 BEC confined in the $x$-$y$ plane 
under a uniform magnetic field applied in the $z$ direction.
For simplicity, we neglect the confining potential in the $x$ and $y$
directions. 
The mean-field kinetic energy and Zeeman energy are given by 
\begin{align}
 E_{\rm kin} &= \int d \bm{r} \sum_{m=-1}^1 \Psi_m^*(\bm{r})
\left(
-\frac{\hbar^2}{2M}\nabla^2\right) \Psi_m(\bm{r}),
\label{eq:E_kin}\\
E_q & =\int d \bm{r} \sum_{m=-1}^1 
q m^2  |\Psi_m(\bm{r})|^2,
\label{eq:E_q}
\end{align}
where $\Psi_m(\bm{r})$ is the condensate wave function for the atoms
in the magnetic sublevel $m$, $M$ is an atomic mass,
and $q$ is the quadratic Zeeman energy per atom.
Here, we neglected the linear Zeeman term because the linear
Zeeman effect merely induces the Larmor precession of atomic spins and
can be eliminated in the rotating frame of reference.
The quadratic Zeeman energy is tunable by means of a
linearly polarized microwave field 
and can take both positive and negative values~\cite{Gerbier,Guzman2011}.

The interatomic interaction energy is given by
\begin{align}
 E_{\rm int} = \frac12\int d\bm{r}\left[
c_0n_{\rm tot}(\bm{r})^2 + c_1|\bm{f}(\bm{r})|^2
\right],
\end{align}
where the number density and the spin density (local magnetization) are
given by
\begin{align}
 n_{\rm tot}(\bm{r}) &=  \sum_{m=-1}^1|\Psi_m(\bm{r})|^2,
\label{eq:n_tot} \\
f_\nu(\bm{r}) &=
 \sum_{m,n=-1}^1\Psi^*_m(\bm{r})(F_\nu)_{mn}\Psi_n(\bm{r}), 
\label{eq:f_mu}
\end{align}
respectively. 
Here, $\nu=x,y,z$ and, $F_{x,y,z}$ are the spin-1 matrices. 
The interaction coefficients are given by
$c_0=4\pi\hbar^2(2a_2+a_0)/(3M)$ and  $c_1=4\pi\hbar^2(a_2-a_0)/(3M)$,
where $a_S$ is the $s$-wave scattering lengths of two colliding atoms with
total spin $S$ channel. 
For the condensate to be stable, $c_0$ needs to be positive. On the
other hand, the sign of $c_1$ determines the magnetism: the condensate is
ferromagnetic (antiferromagnetic or polar) for $c_1<0$ ($c_1>0$). In
this paper, we consider ferromagnetic BECs ($c_1<0$). 

When the quadratic Zeeman energy is weak compared with the ferromagnetic
interaction energy,
the condensate is fully magnetized ($|\bm{f}|=n_{\rm tot}$).
Since the order parameter for a fully-magnetized state in the
direction $(\cos\alpha\sin\beta,\sin\alpha\sin\beta,\cos\beta)$ is
given by~\cite{Kudo10,Kawaguchi12} 
\begin{align}
 \bm\Psi\equiv \begin{pmatrix} \Psi_1 \\ \Psi_0 \\ \Psi_{-1}\end{pmatrix}
 = \sqrt{n_{\rm tot}}e^{i\phi} 
\begin{pmatrix}
 e^{-i\alpha}\cos^2\frac{\beta}{2} \\
 \sqrt{2}\sin\frac{\beta}{2}\cos\frac{\beta}{2} \\
e^{i\alpha}\sin^2\frac{\beta}{2}
\end{pmatrix}, 
\label{eq:Psi.0}
\end{align}
the population in the $m=0$ component becomes maximum at $\beta=\pi/2$,
whereas those in the $m=1$ and $-1$ components become maximum at $\beta=0$
and $\pi$, respectively. 
As seen from Eq.~\eqref{eq:E_q}, the quadratic Zeeman effect enhances the
population in the $m=0$ state for $q>0$ and those in the $m=\pm 1$ states
for $q<0$. 
Hence, the magnetization arises in the $x$-$y$ plane ($\beta=\pi/2$) for
$q>0$, and in the $+z$ or $-z$ direction ($\beta=0$ or $\pi$) for $q<0$. 
The former case corresponds to the $XY$ model and the latter the Ising
model of the conventional ferromagnet. 
Although the magnitude of the spontaneous magnetization becomes smaller
when the quadratic Zeeman energy is positive and comparable to the
ferromagnetic interaction, 
the magnetization direction is still confined in the $x$-$y$ plane.
Since we are interested in vortex-antivortex annihilation, we consider $q>0$
below. 

We first consider a single vortex 
and write its wave function in the polar coordinate 
whose origin is the center of the vortex core: $\bm{\Psi}(r,\varphi)$.
We take $\phi=\sigma_\phi\varphi$ and $\alpha=\sigma_\alpha\varphi$ in
Eq.~\eqref{eq:Psi.0}, 
where $\sigma_\phi$ and $\sigma_\alpha$ are integers.
For a symmetric vortex, $\beta$ is a function of
$r$ and independent of $\varphi$.
At a distance from the vortex core, $\beta=\pi/2$ 
as discussed in the above.
The wave function outside of the core is approximately written as
\begin{align}
 \bm{\Psi}=\frac{\sqrt{n_{\rm tot}}}{2}
 e^{i\sigma_\phi\varphi}
\begin{pmatrix}
 e^{-i\sigma_\alpha\varphi} \\ \sqrt{2} \\ e^{i\sigma_\alpha\varphi}
\end{pmatrix}.
\label{eq:Psi.1}
\end{align}
Here, $\sigma_\phi$ and $\sigma_\alpha$ determine
the directions of mass flow and spin flow around the vortex,
respectively. 
The superfluid velocity (mass flow) and the spin superfluid velocity
(spin flow) of $z$ component are written for a homogeneous $n_{\rm tot}$
as 
\begin{align}
 \bm v_{\rm mass} &= \frac{\hbar}{2Mi}\sum_{m=-1}^1
\left[\Psi_m^*(\bm\nabla\Psi_m)-(\bm\nabla\Psi_m^*)\Psi_m \right]/
n_{\rm tot}, 
\label{eq:vmass} \\
\bm v_{\rm spin}^z &= \frac{\hbar}{2Mi}\sum_{m=-1}^1
(F_z)_{mn}
\left[\Psi_m^*(\bm\nabla\Psi_n)-(\bm\nabla\Psi_m^*)\Psi_n \right]/
n_{\rm tot}, 
\label{eq:vspin}
\end{align}
respectively. Substituting Eq.~\eqref{eq:Psi.1} into
Eqs.~\eqref{eq:vmass} and \eqref{eq:vspin}, we see that the directions
of mass and spin flows depend on $\sigma_\phi$ and $\sigma_\alpha$,
respectively, as
$\bm{v}_{\rm mass}=\sigma_\phi(\hbar/M)\nabla\varphi$ and 
$\bm{v}^z_{\rm spin}=-\sigma_\alpha(\hbar/2M)\nabla\varphi$.

The combination of $\sigma_\phi$ and $\sigma_\alpha$ also determines the
vortex core structure. Though we use $\beta=\pi/2$ in
Eq.~\eqref{eq:Psi.1}, $\beta$ 
changes around the center of the vortex so as to remove the singularity
of the order parameter.
When $\sigma_\phi=\sigma_\alpha$, the $m=1$ component is independent of
$\varphi$ and only this component remains at $r=0$. 
In this case, $\beta$ takes $0$ at $r=0$, which means
the magnetization at the center is in the $+z$ direction
for $\sigma_\phi=\sigma_\alpha$.  
Similarly, for vortices with $\sigma_\phi=-\sigma_\alpha$, 
magnetization is in the $-z$ direction at the center.
When $\sigma_\phi=0$ and $\sigma_\alpha\neq 0$, $\varphi$-dependent
components cannot vanish in a fully-magnetized state.
Thus, magnetization vanishes at the center for $\sigma_\phi=0$.
On the other hand, when $\sigma_\phi\neq 0$ and $\sigma_\alpha=0$, all
three components should vanish at the center.
In the following, we consider only the elementary vortices that are
stable against splitting, that is, $\sigma_\phi=0,\pm 1$ and
$\sigma_\alpha=\pm 1$.
The vortex of $\sigma_\phi=0$ has no mass flow around itself and its
core is not magnetized. Such a vortex is called polar-core vortex
(PCV). 
When $\sigma_\phi=\pm 1$, the vortex 
has mass flow around its core and its core is fully magnetized as well as
the outside. Such a vortex is called Mermin-Ho vortex (MHV).
Considering the combination of $\sigma_\phi$ and $\sigma_\alpha$, 
we notice that there are two kinds of PCVs 
[$(\sigma_\phi,\sigma_\alpha)=(0,\pm1)$]
and four kinds of MHVs
[$(\sigma_\phi,\sigma_\alpha)=(\pm1,\pm1)$].

A vortex-antivortex pair is defined so that they can be pair-annihilated. 
For the case of a single-component BEC, two
vortices with winding numbers
with the opposite signs are a vortex-antivortex pair. In the present case, 
i.e., a multi-component BEC, when two vortices have winding numbers with
the opposite signs in all components, they can be annihilated as a
vortex-antivortex pair. 
For the vortex expressed by Eq.~\eqref{eq:Psi.1}, 
the $m=1$, $0$, and $-1$ components have the winding numbers 
$\sigma_\phi-\sigma_\alpha$, $\sigma_\phi$, and 
$\sigma_\phi+\sigma_\alpha$, respectively. 
Thus, its antivortex is obtained by changing the signs of
both $\sigma_\phi$ and $\sigma_\alpha$. 
In other words, vortices with
$(\sigma_\phi, \sigma_\alpha)$ and $(-\sigma_\phi,-\sigma_\alpha)$ are
a vortex-antivortex pair. 

In this paper, we consider only MHVs, which are useful to investigate
the effect of superfluid flow.
Using Eq.~\eqref{eq:Psi.1}, we estimate the kinetic energy of a single
vortex as
\begin{align}
 E_{\rm s} &= \frac{\hbar^2}{2M}\int d^2r\sum_m
 (\nabla\Psi_m^*)\cdot(\nabla\Psi_m) 
\nonumber \\
&\simeq \frac{\hbar^2}{2M}\int_{R_c}^R rdr\int_0^{2\pi}d\varphi\sum_m
 (\nabla\Psi_m^*)\cdot(\nabla\Psi_m)
\nonumber \\
&= C\mathcal{N}_1\log\left(\frac{R}{R_c}\right),
\label{eq:E_s}
\end{align} 
where $C=\pi\hbar^2n_{\rm tot}/M$, and $R$ and $R_c$ are the vortex size
(radius) and the radius of the vortex core, respectively. 
Although $R$ is equal to the system size for a single vortex,
it is the distance beyond which the field around the vortex is shielded
if there are other vortices.
$\mathcal{N}_1$ depends on $\sigma_\phi$ and
$\sigma_\alpha$. 
Since the portions of the number densities for $m=1,0,-1$ are
$\frac14,\frac12,\frac14$, respectively,
\begin{align}
 \mathcal{N}_1=\frac{\nu_1^2}4  + \frac{\nu_0^2}2  + \frac{\nu_{-1}^2}4, 
\label{eq:N1}
\end{align}
where $\nu_1=\sigma_\phi-\sigma_\alpha$, $\nu_0=\sigma_\phi$, and 
$\nu_{-1}=\sigma_\phi+\sigma_\alpha$.

\begin{figure}[tb]
\includegraphics[width=8.5cm,clip]{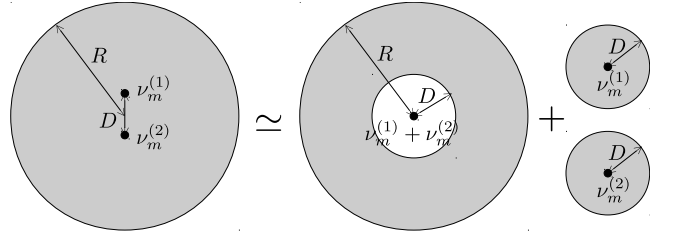}
\caption{
Evaluation of the energy $E_{\rm pair}$ of two vortices with
 vorticities $\nu_m^{(1)}$ and $\nu_m^{(2)}$ separated by a distance $D$.
}   
\label{fig:Epair} 
\end{figure}

The interaction energy between two vortices at a distance of $D$ is given by
$V(D)=E_{\rm pair}(D)-E_1-E_2$, where $E_{\rm pair}$ is the energy of
two vortices 
separated by a distance $D$, 
$E_1$ and $E_2$ are the energies of single vortices with
vorticities $\nu^{(1)}_m$ and $\nu^{(2)}_m$, respectively. 
The pair energy $E_{\rm pair}$ is approximately given by
the sum of contributions from two regions (see Fig.~\ref{fig:Epair}).
In the region of $D<r<R$, the contribution is evaluated 
for a single (composite) vortex with vorticity
$\nu^{(1)}_m+\nu^{(2)}_m$.
In the region of $r<D$, the contribution is just 
the sum of $E_1$ and $E_2$.   
Then, the interaction energy is approximated by
\begin{align}
 V(D)&=C\mathcal{N}_2\log\left(\frac{R}{D}\right),
\label{eq:VD} \\
\mathcal{N}_2 &= 
\frac{\nu^{(1)}_1\nu^{(2)}_1}2  
+ \nu^{(1)}_0\nu^{(2)}_0 + \frac{\nu^{(1)}_{-1}\nu^{(2)}_{-1}}2.
\label{eq:N2}
\end{align} 
The derivative of $V(D)$ gives the force between the vortex pair,
\begin{align}
 F_{\rm pair}=-\frac{dV(D)}{dD}=\frac{C\mathcal{N}_2}{D},
\label{eq:F_p}
\end{align}
which is an attractive force between a vortex-antivortex pair.
Note that pair annihilation occurs only between a vortex pair in
which both $\sigma_\phi$ and $\sigma_\alpha$ have the opposite
signs, although the force is attractive ($\mathcal{N}_2<0$)
between vortices with the opposite signs of $\sigma_\phi$ even if they
have the same sign of $\sigma_\alpha$.

\subsection{Coarsening dynamics}
\label{sec:Ed}

We assume that the attractive force 
between a vortex-antivortex pair
is balanced with a friction 
(resistive) force $F_{\rm fric}$ when the pair vortices move toward each other.
The friction causes energy dissipation.
When a vortex moves at speed $u$, the energy dissipation rate is written
as
\begin{align}
 \frac{dE}{dt}=-uF_{\rm fric}.
\label{eq:dEdt.1}
\end{align}

The dynamics of a spinor BEC is well described with the
time-dependent multi-component Gross-Pitaevskii (GP) equation, and we
phenomenologically introduce an energy dissipation into the GP
equation~\cite{Tsubota,Kawaguchi12}: 
\begin{align}
 (i-\Gamma) &\hbar\frac{\partial}{\partial t}\Psi_m(\bm{r},t)
\nonumber\\
&=\left[-\frac{\hbar^2}{2M}\nabla^2-\mu(t)
 +qm^2+c_0n_{\rm tot}(\bm{r},t)\right]\Psi_m(\bm{r},t)
\nonumber\\
&\quad +c_1\sum_{n=-1}^1\sum_{\nu=x,y,z}f_\nu(\bm{r},t)(F_\nu)_{mn}
\Psi_n(\bm{r},t),
\label{eq:GP}
\end{align} 
where $\Gamma$ expresses energy dissipation.

In order to discuss the energy dissipation that is caused by the friction
force, we employ the hydrodynamic equation, which 
is derived in the low-energy limit~\cite{Kudo11,Kudo13}.
In this limit, the BEC is fully magnetized, i.e.,
$|\bm{f}|=n_{\rm tot}$, and
the physical quantities that describe the dynamics
of ferromagnetic BECs are the normalized spin vector 
\begin{align}
 \hat{\bm f} \equiv \frac{\bm f}{n_{\rm tot}},
\end{align}
and the superfluid velocity ${\bm v}_{\rm mass}$.
The equations of motion for them are derived
straightforwardly from the GP
equation~\eqref{eq:GP}~\cite{Kudo10,Kudo11,Kudo13}, 
and the resulting equations of motion are written as
\begin{subequations}
\begin{align}
 \frac{\partial \hat{\bm{f}}}{\partial t}
 &= \frac{1}{1+\Gamma^2}\left[
 \frac{1}{\hbar}\hat{\bm{f}} \times \bm{B}_{\rm eff}
 - (\bm{v}_{\rm mass}\cdot\nabla)\hat{\bm{f}}
 \right]
\nonumber \\
 &\quad - \frac{\Gamma}{1+\Gamma^2}\hat{\bm{f}} \times\left[
 \frac{1}{\hbar}\hat{\bm{f}} \times \bm{B}_{\rm eff}
 - (\bm{v}_{\rm mass}\cdot\nabla)\hat{\bm{f}}
 \right],
 \label{eq:extLLG}
 \\
 \bm{B}_{\rm eff} &= \frac{\hbar^2}{2M}\nabla^2\hat{\bm{f}}
 - q\hat{f}_z\hat{z},
 \label{eq:B_eff}
\\
 M\frac{\partial}{\partial t} \bm{v}_{\rm mass} 
 &= \frac{\hbar}{2n_{\rm tot}\Gamma}\nabla\left[ 
 \nabla\cdot(n_{\rm tot}\bm{v}_{\rm mass})  \right]
\nonumber \\
&\quad
 + \hbar (\nabla\hat{\bm{f}})\cdot\left(
 \hat{\bm{f}}\times\frac{\partial\hat{\bm{f}}}{\partial t}
 \right).
 \label{eq:dvdt}
 \end{align}
\label{eq:HDE}
\end{subequations}
Here, we assumed a uniform number density: $\nabla n_{\rm tot}=0$.

The kinetic energy in this formulation is written as
\begin{align}
 E_{\rm kin} &= \frac{\hbar^2n_{\rm tot}}{4M}\int d\bm{r}\left[
(\nabla\hat{f}_x)^2 + (\nabla\hat{f}_y)^2 + (\nabla\hat{f}_z)^2
\right] \nonumber\\
&\quad 
+ \frac{Mn_{\rm tot}}{2}\int d\bm{r}\; \bm{v}_{\rm mass}^2.
\label{eq:E_kin.h}
\end{align}
We divide the energy dissipation into two parts, which is written as
\begin{align}
 \frac{dE}{dt} &= \frac{dE_{\rm mag}}{dt} + \frac{dE_{\rm flow}}{dt},
\\
 \frac{dE_{\rm mag}}{dt} &= \int d\bm{r}\left(
\frac{\delta E}{\delta\hat{f}_x}\frac{\partial\hat{f}_x}{\partial t}
+ \frac{\delta E}{\delta\hat{f}_y}\frac{\partial\hat{f}_y}{\partial t}
+ \frac{\delta E}{\delta\hat{f}_z}\frac{\partial\hat{f}_z}{\partial t}
\right), 
\label{eq:dE_mag}\\
\frac{dE_{\rm flow}}{dt} &= \int d\bm{r}\left(
\frac{\delta E}{\delta v_x}\frac{\partial v_x}{\partial t}
+ \frac{\delta E}{\delta v_y}\frac{\partial v_y}{\partial t}
\right),
\label{eq:dE_flow}
\end{align}
where $\bm{v}_{\rm mass}=(v_x,v_y)$.
We assume that a vortex keeps its shape, i.e., the profiles of
$\bm{v}_{\rm mass}$ and $\hat{\bm f}$ around its core, when it moves.
The contribution to $dE_{\rm mag}/dt$ arises from the change in
direction of local magnetization. Since the profiles of
$\bm{v}_{\rm mass}$ and $\hat{\bm f}$ are conserved, 
the coupling between $\bm{v}_{\rm mass}$ and $\hat{\bm f}$ gives no 
contribution to $dE_{\rm mag}/dt$.
Neglecting the energy contributions from the vortex core, 
we only need to consider the hydrodynamic equation in the outside region
of the vortex core.
Then, Eq.~\eqref{eq:HDE} with 
$\hat{f}_z=\partial\hat{f}_z/\partial t=\nabla\hat{f}_z=0$ leads to
\begin{subequations}
\begin{align}
 \frac{\partial\hat{f}_x}{\partial t} &= 
- \frac{\Gamma}{1+\Gamma^2}\frac{\hbar}{2M}\left[
\hat{f}_x (\hat{\bm f}\cdot\nabla^2\hat{\bm f}) - \nabla^2\hat{f}_x
\right], \label{eq:dfxdt}\\
 \frac{\partial\hat{f}_y}{\partial t} &= 
- \frac{\Gamma}{1+\Gamma^2}\frac{\hbar}{2M}\left[
\hat{f}_y (\hat{\bm f}\cdot\nabla^2\hat{\bm f}) - \nabla^2\hat{f}_y
\right], \label{eq:dfydt}\\
 \frac{\partial\bm{v}_{\rm mass}}{\partial t} &= \frac{\hbar}{2M\Gamma}
\nabla(\nabla\cdot\bm{v}_{\rm mass}),\label{eq:dvdt.1}
\end{align}
\label{eq:redHDE}%
\end{subequations}
where the coupling terms with 
$(\bm{v}_{\rm mass}\cdot\nabla)\hat{\bm f}$ are dropped. 
Since we take $\hat{f}_z=0$, $\hat{f}_x=\cos\alpha$ and
$\hat{f}_y=\sin\alpha$.
From Eqs.~\eqref{eq:dfxdt} and \eqref{eq:dfydt}, we have
\begin{align}
 \frac{\partial\alpha}{\partial t} 
=\hat{f}_x\frac{\partial\hat{f}_y}{\partial t}
- \hat{f}_y\frac{\partial\hat{f}_x}{\partial t} 
=  \frac{\Gamma}{1+\Gamma^2}\frac{\hbar}{2M}\nabla^2\alpha.
\label{eq:dadt}
\end{align}
Substituting Eqs.~\eqref{eq:dfxdt} and \eqref{eq:dfydt} into
Eq.~\eqref{eq:dE_mag} gives
\begin{align}
 \frac{dE_{\rm mag}}{dt} &= -\frac{\hbar^2n_{\rm tot}}{2M}
\int d\bm{r}\left(
(\nabla^2\hat{f}_x)\frac{\partial\hat{f}_x}{\partial t}
+ (\nabla^2\hat{f}_y)\frac{\partial\hat{f}_y}{\partial t}
\right) \nonumber\\
&= -\frac{\hbar n_{\rm tot}(1+\Gamma^2)}{\Gamma}
\int d\bm{r}
\left[ \left(\frac{\partial\hat{f}_x}{\partial t}\right)^2
+ \left(\frac{\partial\hat{f}_y}{\partial t}\right)^2 \right]
\nonumber\\
&= -\frac{\hbar n_{\rm tot}(1+\Gamma^2)}{\Gamma}
\int d\bm{r} \left(\frac{\partial\alpha}{\partial t}\right)^2,
\label{eq:dE_mag.1}
\end{align}
where we used $\hat{f}_x^2+\hat{f}_y^2=1$.
Substituting Eq.~\eqref{eq:Psi.1} into Eq.~\eqref{eq:vmass}, we have
\begin{align}
 \bm{v}_{\rm mass} = \frac{\hbar}{M}\nabla\phi,
\label{eq:vmass.1}
\end{align}
where $\phi=\sigma_\phi\varphi$. Equations.~\eqref{eq:dvdt.1} and
\eqref{eq:vmass.1} lead to
\begin{align}
 \frac{\partial\phi}{\partial t}=\frac{\hbar}{2M\Gamma}\nabla^2\phi.
\label{eq:dpdt}
\end{align}
Using Eqs.~\eqref{eq:dvdt.1}, \eqref{eq:vmass.1}, and \eqref{eq:dpdt}, 
we rewrite Eq.~\eqref{eq:dE_flow} as
\begin{align}
  \frac{dE_{\rm flow}}{dt} &= Mn_{\rm tot}\int d\bm{r}\;
\bm{v}_{\rm mass}\cdot\frac{\partial\bm{v}_{\rm mass}}{\partial t}
\nonumber\\
&= -2\hbar n_{\rm tot}\Gamma\int d\bm{r} 
\left(\frac{\partial\phi}{\partial t}\right)^2.
\label{eq:dE_flow.1}
\end{align}
Suppose that a vortex keeps its shape when it
moves in the $x$ direction at speed $u$: 
$\alpha(\bm{r})=f(x-ut,y)$ and $\phi(\bm{r})=g(x-ut,y)$,
where $f$ and $g$ are functions expressing their profiles. 
Then, 
$(\partial\alpha/\partial t)^2=u^2(\partial\alpha/\partial x)^2$ and
$(\partial\phi/\partial t)^2=u^2(\partial\phi/\partial x)^2$.
Similarly, for a vortex moving in the $y$ direction,
$(\partial\alpha/\partial t)^2=u^2(\partial\alpha/\partial y)^2$ and
$(\partial\phi/\partial t)^2=u^2(\partial\phi/\partial y)^2$.
The averages of them result in
\begin{align}
 \left(\frac{\partial\alpha}{\partial t}\right)^2 
= \frac{u^2}{2}(\nabla\alpha)^2,
\quad
 \left(\frac{\partial\phi}{\partial t}\right)^2 
= \frac{u^2}{2}(\nabla\phi)^2.
\label{eq:dadtdpdt}
\end{align}
Combining Eqs.~\eqref{eq:dE_mag.1} and \eqref{eq:dE_flow.1} and using 
Eq.~\eqref{eq:dadtdpdt}, we estimate the energy dissipation as
\begin{align}
 \frac{dE}{dt} &= 
-\frac{\hbar n_{\rm tot}}{\Gamma}\int d\bm{r}\left[
(1+\Gamma^2)\left(\frac{\partial\alpha}{\partial t}\right)^2
+ 2\Gamma^2\left(\frac{\partial\phi}{\partial t}\right)^2
\right]
\nonumber\\
&=-\frac{2M}{\hbar\Gamma}
C\mathcal{N}_{\rm fric}\log\left(\frac{R}{R_c}\right)u^2.
\label{eq:dEdt.2}
\end{align}
Here,
\begin{align}
 \mathcal{N}_{\rm fric} = (1+\Gamma^2)\mathcal{N}_1-\sigma_\phi^2,
\label{eq:Nfric}
\end{align}
where we used $\alpha=\sigma_\alpha\varphi$.
The friction force is estimated by comparing
Eqs.~\eqref{eq:dEdt.1} and \eqref{eq:dEdt.2}:
\begin{align}
 F_{\rm fric}=\frac{2M}{\hbar\Gamma}
C\mathcal{N}_{\rm fric}\log\left(\frac{R}{R_c}\right)u.
\label{eq:F_f}
\end{align}

In order to investigate the growth of characteristic domain size $\xi$, 
we apply the discussion of the coarsening dynamics in the 2D $XY$
model~\cite{Bray,Pargellis,Yurke,Rutenberg}, where we expect 
$\xi\sim R\sim D$ and $u=d\xi/dt$.
Equating the characteristic force between a vortex pair 
$F_{\rm pair}\propto\mathcal{N}_2/\xi$ with the characteristic friction
force 
$F_{\rm fric}\propto(\mathcal{N}_{\rm fric}/\Gamma)\log(\xi/R_c)d\xi/dt$,
and rearranging terms, we have
\begin{align}
 \xi\log\left(\frac{\xi}{R_c}\right)\frac{d\xi}{dt}=A,
\label{eq:dxdt.1}
\end{align} 
where $A$ is a constant that depends on the dissipation rate and the
characteristics of vortices as
$A\propto -\Gamma\mathcal{N}_2/\mathcal{N}_{\rm fric}$.
Integrating of Eq.~\eqref{eq:dxdt.1} gives
\begin{align}
 \xi^2\left[ \log\left(\frac{\xi}{R_c}\right)-\frac12 \right]
=2A(t-t_0),
\label{eq:xi.1}
\end{align}
where $t_0$ is an integration constant.
Employing the vortex density 
$\rho=1/\xi^2$ and 
the maximum vortex density $\rho_c=1/R_c^2$, 
we rewrite Eq.~\eqref{eq:xi.1} as
\begin{align}
 t-t_0=\frac{1}{4A}\frac{\log(\rho_c/\rho)-1}{\rho}.
\label{eq:rho.1}
\end{align}
The number of vortices in ferromagnetic BECs is expected to yield
Eq.~\eqref{eq:rho.1}, 
which is the same as the growth law for the $XY$ model~\cite{Yurke}.
The difference between the $XY$ model and
ferromagnetic BECs is contained in factor $A$, which includes the
information about vortices ($\mathcal{N}_{\rm fric}$ and
$\mathcal{N}_2$). 
Actually, the hydrodynamic equation with $\bm{v}_{\rm mass}=0$
corresponds to the $XY$ model when $\hat{f}_z\simeq 0$ 
(namely, in a positive-$q$ case).
If $\bm{v}_{\rm mass}=0$, we just drop the $E_{\rm flow}$ terms in the
above discussion, and
then obtain the same equation as Eq.~\eqref{eq:rho.1}, although the
factor $A$ is different from that of the above case.

\section{Numerical simulations}
\label{sec:simu}

We perform numerical simulations by means of the dissipative GP
equation~\eqref{eq:GP} 
and the hydrodynamic equation~\eqref{eq:HDE}.
The advantage of the hydrodynamic equation is that 
the superfluid velocity $\bm{v}_{\rm mass}$
can be eliminated easily in simulations,
which enables us to investigate what effects the superfluid flow has on
the coarsening dynamics. 
Note that MHVs introduced in Sec.~\ref{sec:vortex} have both mass flow
and spin flow around their cores.　
However, if we take $\bm{v}_{\rm mass}=0$ in simulations, 
the degrees of freedom of mass flow are eliminated. Then, the 
hydrodynamic equation reduces to the equation of motion of
magnetizations, and what we call MHV in the discussion
below becomes merely a spin vortex around which only the spin current 
circulates. In such a case, the index $\sigma_\phi$
is meaningless, and there exists only one combination of a
vortex-antivortex pair; $\sigma_\alpha=1$ and $-1$. 

In the simulations, the mass of an atom is given by a typical value for
a spin-1 $^{87}$Rb atom: $M=1.44\times 10^{-25}$ kg. 
The total number density is taken as 
$n_{\rm tot}=\sqrt{2\pi d^2}n_{\rm 3D}$ with 
$n_{\rm 3D}=2.3\times 10^{14}$ cm$^{-3}$ and $d=1$ $\mu$m. 
The quadratic Zeeman energy is set to be $q/h=10$ Hz.
The dissipation rate is given by a typical value $\Gamma=0.03$.
Especially in the GP simulation, the system is in quasi-two dimensions: 
The wave function in the normal direction to the 2D plane is
approximated by a 
Gaussian with width $d$. Interaction parameters are taken as 
$c_0n_{\rm 3D}/h = 1.3$ kHz and 
$c_1n_{\rm 3D}=-59$ Hz. 
The value of $c_1$ that we take here is ten times
larger than a typical value of a spin-1 $^{87}$Rb atom, which
prevents the production of PCVs.

\begin{figure}[tb]
\includegraphics[width=8.5cm,clip]{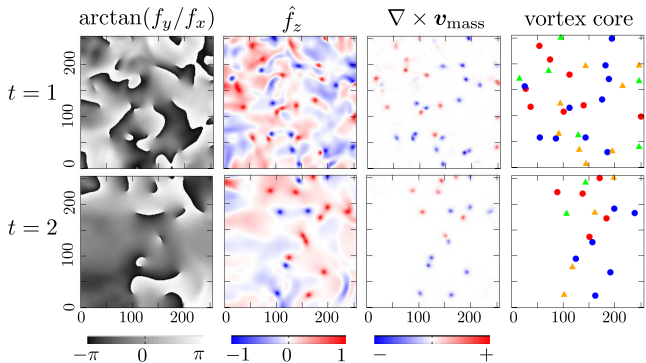}
\caption{(Color online) Snapshots of the transverse (arctan$(f_y/f_x)$)
 and longitudinal ($\hat{f}_z$) magnetizations, the vorticity of 
 mass flow ($\nabla\times\bm{v}_{\rm mass}$), and the positions
 of vortex cores at time $t=1$ s (top) and $t=2$ s (bottom) are 
 shown in the $x$-$y$ plane, 
 which are simulated by hydrodynamic simulations. 
 The size of the snapshot is $256$ $\mu$m on each side.
 The color of vortex core represents the directions of
 mass and spin flows $(\sigma_\phi=\pm1,\sigma_\alpha=\pm1)$. 
 Vortices that make annihilation pairs have the same symbol shape: 
 red ($+,+$) and blue ($-,-$) circles, and
 green ($+,-$) and orange ($-,+$) triangles.
}   
\label{fig:snap} 
\end{figure}

\begin{table}[tb]
\begin{tabular}{|c|c|c|}
\hline
$(\sigma_\phi,\sigma_\alpha)$ & $\hat{f}_z$ & 
$\nabla\times\bm{v}_{\rm mass}$\\
\hline
$(+,+)$ & $+$ & $+$ \\
\hline
$(-,-)$ & $+$ & $-$ \\
\hline
$(+,-)$ & $-$ & $+$ \\
\hline
$(-,+)$ & $-$ & $-$ \\
\hline
\end{tabular} 
\caption{Signs of $\hat{f}_z$ and
$\nabla\times\bm{v}_{\rm mass}$ at  vortex cores with 
$(\sigma_\phi=\pm1,\sigma_\alpha=\pm1)$.}
\label{tab:vortex} 
\end{table}

Initial states are given by randomly located four kinds of MHVs. The
number of vortices of each kind is equal. Open boundary conditions are
imposed on $256$ $\mu$m $\times$ $256$ $\mu$m systems. The total number
of vortices at first are $256$, which implies that the average distance
between vortices is about $16$ $\mu$m.
Snapshots of the transverse and longitudinal magnetizations, 
the vorticity of 
mass flow, and the positions of vortex cores are demonstrated
in Fig.~\ref{fig:snap}. The positions of vortex cores agree with those
of maxima of $\nabla\times\bm{v}_{\rm mass}$. The transverse
magnetization and the vorticity of 
$\bm{v}_{\rm mass}$ are used to
classify vortices into four kinds: $(\sigma_\phi,\sigma_\alpha)=$
($+,+$), ($-,-$), ($+,-$) and ($-,+$).
The combination of $\sigma_\phi$ and $\sigma_\alpha$ is also related to 
the longitudinal magnetization at a vortex core,
positive (negative) $\hat{f}_z$ for $\sigma_\phi=\sigma_\alpha$ 
($\sigma_\phi=-\sigma_\alpha$), as mentioned in Sec.~\ref{sec:vortex}.
The sign of $\nabla\times\bm{v}_{\rm mass}$ is related to the
combination of $\sigma_\alpha$ and $\hat{f}_z$ or simply $\sigma_\phi$.
Table~\ref{tab:vortex} shows the signs of $\hat{f}_z$ and
$\nabla\times\bm{v}_{\rm mass}$ at vortex cores for all the combinations
of $(\sigma_\phi,\sigma_\alpha)$. 
The vortices with ($+,+$) and ($-,-$), which are represented as circles
in Fig.~\ref{fig:snap}, are a vortex-antivortex pair. Those with ($+,-$)
and ($-,+$), which are represented as triangles, are 
another vortex-antivortex pair. 

\begin{figure}[tb]
\includegraphics[width=8cm,clip]{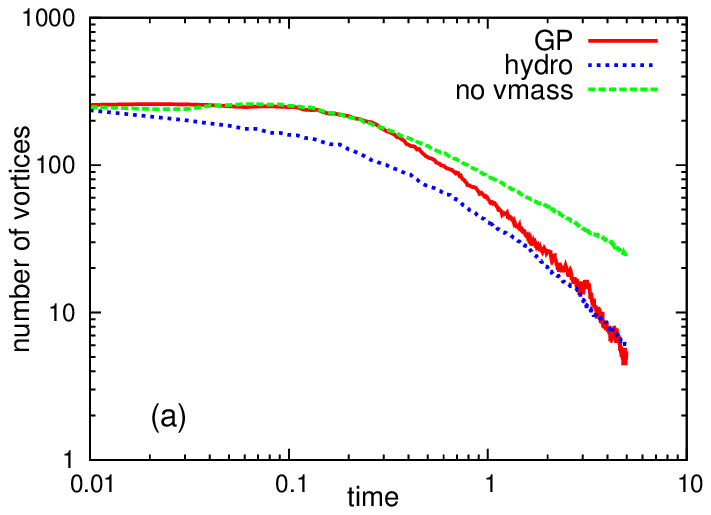}
\includegraphics[width=8cm,clip]{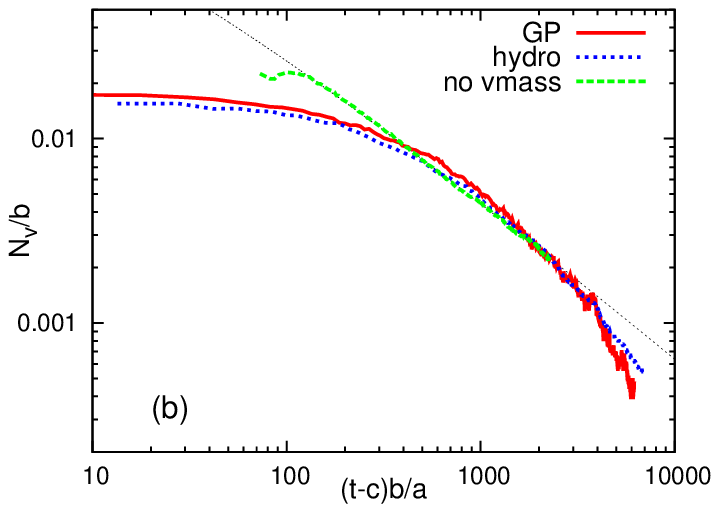}
\caption{(Color online) (a) Time dependence of the number of vortices
 simulated by GP equation (GP), hydrodynamic equation equation
 (hydro), and 
 Eq.~\eqref{eq:HDE} in the absence of $\bm{v}_{\rm mass}$ (no vmass).
Each curve is the average of ten simulations.
(b) The same data replotted to show how they fit to 
 $X=(\log(1/Y)-1)/Y$ (dashed line)
 by means of the scaling Eq.~\eqref{eq:rho.1}.
}    
\label{fig:mhv4} 
\end{figure}

The number of vortices decreases with time as shown in
Fig.~\ref{fig:mhv4}(a). In the case of no superfluid flow, which is
simulated by Eq.~\eqref{eq:HDE} with $\bm{v}_{\rm mass}=0$ at all 
times, the decay is slower than the other simulations.
This suggests that the superfluid flow has an effect to accelerate
the coarsening dynamics.
However, the effect is not very simple, which is suggested in
Fig.~\ref{fig:mhv4}(b). The dashed line represents the scaling,
Eq.~\eqref{eq:rho.1}. The data are fitted to 
\begin{align}
 t=a[\log(b/N_{\rm v})-1]/N_{\rm v}+c,
\end{align}
where $t$ and $N_{\rm v}$ are time and the number of vortices, 
respectively. 
Note that $N_{\rm v}=\rho L^2$, where $L$ is the system size and $L=256$
$\mu$m in the simulations.
For the fitting, the data in the range of 
$N_{\rm v} \ge 20$ are used.
The fitting parameters $a$ and $c$ correspond to 
$L^2/4A$ and $t_0$, respectively. Parameter $b$, which
corresponds to $\rho_cL^2$, is set to be a constant value, 
$b=(256/2.4)^2$. Actually, the core size is estimated to be 
$R_c\simeq 2.4$ $\mu$m in the condensate of spin-$1$ $^{87}$Rb atoms 
for $q/h=10$ Hz.
Since the fitting function is modified to be
$(t-c)b/a=[\log(b/N_{\rm v})-1](b/N_{\rm v})$, 
the data are plotted as 
$X=(t-c)b/a$ and $Y=N_{\rm v}/b$, 
and they are expected to be on the curve
$X=[\log(1/Y)-1]/Y$.
The values of fitting parameters in Fig.~\ref{fig:mhv4}(b) are 
$(a,c)=(8.7,0.24)$ in the GP simulation, 
$(8.1,0.050)$ in the hydrodynamic simulation, and
$(25.4,-0.16)$ 
in the absence of $\bm{v}_{\rm mass}$.
Although the data in the absence of $\bm{v}_{\rm mass}$ fit to the curve
well, those of  GP  and hydrodynamic simulations are very
different from the expected scaling.

It might look strange that 
the curves of the GP and hydrodynamic simulations behave different in
Fig.~\ref{fig:mhv4}~(a), although they are similar in
Fig.~\ref{fig:mhv4}~(b). 
Actually, just the early-time dynamics is different between GP and
hydrodynamic simulations. 
The given initial states, which are unstable, strongly affect the
early-time dynamics. 
After the early time, both the GP and hydrodynamic simulations follow a
common growth law. 
Since the growth law is not just a power law, the rescaled plots in
Fig.~\ref{fig:mhv4}~(b) behave similar even though they look different
in the original plots.

The difference of situations between the GP and hydrodynamic
simulations and the simulation without $\bm{v}_{\rm mass}$ is twofold.
First, the superfluid flow may reduce friction (resistivity) in
ferromagnetic BECs,
which results in the faster decay of the number of vortices
in the GP and hydrodynamic simulations than the simulation without 
$\bm{v}_{\rm mass}$. 
Second, there is only one combination
of vortex-antivortex pair in the absence of superfluid flow, which is
the same situation in the $XY$ model.
In other words, when $\bm{v}_{\rm mass}=0$, $\sigma_\phi$ is
meaningless, and
vortices with the opposite signs
of $\sigma_\alpha$ make a vortex-antivortex pair.
By contrast, there are two 
combinations of vortex-antivortex pairs
[i.e., one is $(\sigma_\phi,\sigma_\alpha)=(+,+)$ and $(-,-)$, and
the other is $(+,-)$ and $(+,-)$]
in the GP and hydrodynamic simulations.

\begin{figure}[tb]
\includegraphics[width=8cm,clip]{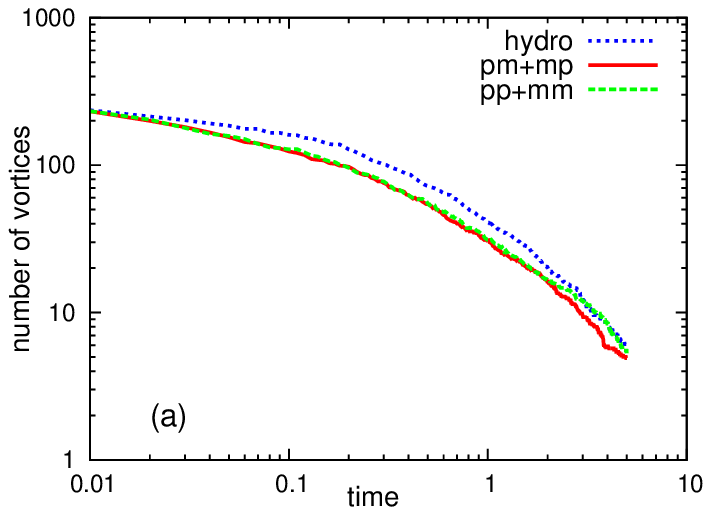}
\includegraphics[width=8cm,clip]{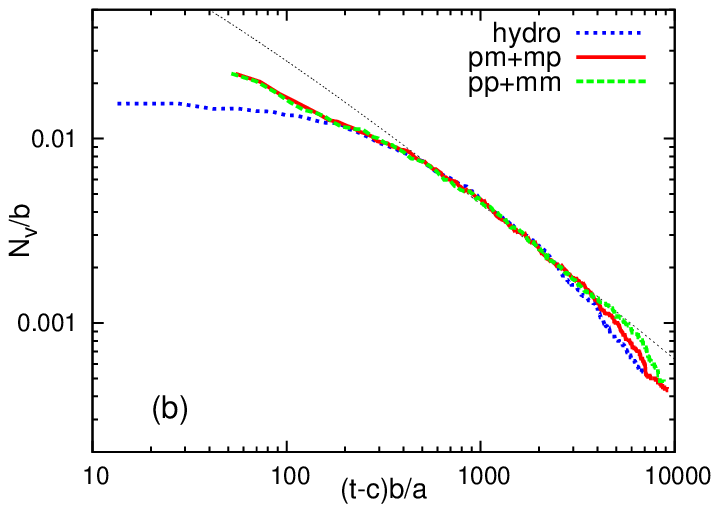}
\caption{(Color online) (a) Time dependence of the number of vortices
 simulated by hydrodynamic equations in the case where 
 there are only two kinds of vortices, 
 $(\sigma_\phi,\sigma_\alpha)=(+,+),(-,-)$
 (labeled with ``pp+mm'') or $(+,-),(-,+)$ (labeled with ``pm+mp''). 
 They decay faster than the case where four kinds of vortices exist,
 whose label is ``hydro'' (the same data in Fig.~\ref{fig:mhv4} (a)).
Each curve is the average of 10 simulations.
(b) The same data replotted to show how they fit to
 $X=(\log(1/Y)-1)/Y$ (dashed line) by means of the scaling
 Eq.~\eqref{eq:rho.1}.}   
\label{fig:mhv2} 
\end{figure}

In order to clarify the reason why the data in the above GP and
hydrodynamic 
simulations do not agree with the expected scaling, we demonstrate the
simulations in special cases where MHVs are limited to two kinds that 
can make a vortex-antivortex pair.
MHVs of $(\sigma_\phi,\sigma_\alpha)=(+,+)$ are pair-annihilated 
with those of
$(-,-)$ but not with the other kinds, $(+,-)$ or $(-,+)$.
If there are only MHVs of $(+,+)$ and $(-,-)$, there is only one
combination of annihilation pairs, which is the same situation as the
simulation in the absence of superfluid flow.
Then, we can see pure effects of $\bm{v}_{\rm mass}$ on the coarsening
dynamics.  The simulations with MHVs of $(+,-)$ and $(-,+)$ also give
the same situation.
In Fig.~\ref{fig:mhv2}(a),
the number of vortices decays slightly faster than the hydrodynamic
simulations, and thus, much faster than the simulation
in the absence of $\bm{v}_{\rm mass}$. 
This fact implies that the coarsening dynamics is accelerated by
superfluid flow. 
On the other hand, Fig.~\ref{fig:mhv2}(b) illustrates better fitting for the 
two-kind-vortex data (labeled by ``pp+mm'' and ``pm+mp'') than the
four-kind-vortex data (labeled by ``hydro'').  
The values of fitting parameters in Fig.~\ref{fig:mhv2}(b) are 
$(a,c)=(6.5,-0.030)$ for ``pp+mm'', and $(6.2,-0.030)$ for ``pm+mp''.
This result indicates that  
the scaling that describes the time dependence
of vortex density is different from the expected one,
Eq~\eqref{eq:rho.1}, when there are several 
combinations of vortex-antivortex pairs. 

\section{Revised Growth Law}
\label{sec:rev}

We here consider revising the scaling, and hence the growth law, 
for the case where there are two combinations
(groups) of vortex-antivortex pairs. 
Vortices belonging to different groups cannot cause annihilation with each
other. When the groups of vortex density $\rho_1$ and $\rho_2$ are
mixed and coexist in the same space, we rewrite the total vortex density 
$\rho_{\rm tot}=\rho_1+\rho_2$  as
\begin{align}
 \rho_{\rm tot}= 2\tilde{\rho}-2\rho_0,
\label{eq:rho_t}
\end{align}
where $\tilde{\rho}$ represents a typical vortex density of a group and
is supposed to obey the original scaling Eq.~\eqref{eq:rho.1}, and
$2\rho_0$ corresponds to the difference between the expected and actual
vortex densities.
Suppose $\rho_1=\tilde{\rho}$ and $\rho_2=\tilde{\rho}-2\rho_0$, where 
$\rho_0>0$.
This implies that the vortices with density $\rho_1$, which are in the
majority, dominate the coarsening dynamics.
The difference between them $\rho_1-\rho_2=2\rho_0$ is almost
independent of time, because the decay rate of vortex density is
similar to each other if $\rho_1\simeq\rho_2$.
Even if $\rho_1=\rho_2$ in the initial state, difference in vortex
density often arises in a early time, when the vortex density is too
high to obey the scaling.
The difference is small, however, it is the key in the revised scaling.

\begin{figure}[tb]
\includegraphics[width=8cm,clip]{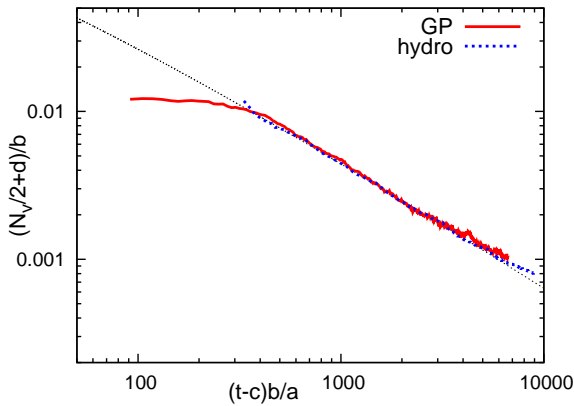}
\caption{(Color online) The same data of the GP and hydrodynamic
 simulations as Fig.~\ref{fig:mhv4} replotted to show how they fit to
 $X=(\log(1/Y)-1)/Y$ (dashed line)
by means of the revised scaling Eq.~\eqref{eq:rho.2}.   
}
\label{fig:mod} 
\end{figure}

Substituting $\tilde{\rho}$ of Eq.~\eqref{eq:rho_t} into 
$\rho$ of Eq.~\eqref{eq:rho.1}, we obtain a
revised equation
\begin{align}
 t-t_0=\frac{1}{4A}
\frac{\log[\rho_c/(\rho_{\rm tot}/2+\rho_0)]-1}{\rho_{\rm tot}/2+\rho_0}.
\label{eq:rho.2}
\end{align}
The same data of the GP and hydrodynamic simulations as that of
Fig.~\ref{fig:mhv4}, which are fitted to Eq.~\eqref{eq:rho.2}, are shown
in Fig.~\ref{fig:mod}. 
The data are actually fitted to 
\begin{align}
 t=a\{\log[b/(N_{\rm v}/2+d)]-1\}/(N_{\rm v}/2+d)+c
\end{align}
with $b=(256/2.4)^2$.  
The fitting parameters $a$ and $c$ correspond to 
$L^2/4A$ and $t_0$, respectively, and $d$ corresponds to $\rho_0L^2$. 
Since the fitting function is modified to be
$(t-c)b/a=\{\log[b/(N_{\rm v}/2+d)]-1\}[b/(N_{\rm v}/2+d)]$, 
the data are plotted as 
$X=(t-c)b/a$ and $Y=(N_{\rm v}/2+d)/b$, 
and they are expected to be on the curve 
$X=[\log(1/Y)-1]/Y$.
The values of fitting parameters in Fig.~\ref{fig:mod} are 
$(a,c,d)=(8.6,-0.059,9.3)$ and $(6.4,-0.19,6.0)$ 
in the GP and hydrodynamic simulations, respectively.
The data are in good agreement with the revised scaling.

\section{Discussions and Conclusions}
\label{sec:disc}

We here discuss fitting parameters quantitatively.
The the fitting parameter 
$a$, which corresponds to 
$L^2/4A\propto\mathcal{N}_{\rm fric}/\mathcal{N}_2$, 
is different between the presence and absence of superfluid flow.
From Eqs.~\eqref{eq:N1}, \eqref{eq:N2}, and \eqref{eq:Nfric}, 
$\mathcal{N}_{\rm fric}/\mathcal{N}_2=-(1+3\Gamma^2)/6$ in the presence
of superfluid flow.
When there is no superfluid flow, 
$\mathcal{N}_{\rm fric}/\mathcal{N}_2
=(1+\Gamma^2)\mathcal{N}'_1/\mathcal{N}'_2$, 
where $\mathcal{N}'_1$ and $\mathcal{N}'_2$ are given by 
the same winding numbers as the 2D $XY$ model, and thus,
$\mathcal{N}'_1/\mathcal{N}'_2=-1/2$.
This means
$\mathcal{N}_{\rm fric}/\mathcal{N}_2=-(1+\Gamma^2)/2$
in the absence of superfluid flow. 
Thus, the value of $a$ in the simulation in the absence of
superfluid flow should be about three times larger than that in the GP and
hydrodynamic simulations.
Actually, in Fig.~\ref{fig:mhv4} (b), $a=8.7$ and $8.1$ for the GP and
hydrodynamic simulations, respectively, and they are about $1/3$ of
$a=25.4$ for the simulation in the absence of superfluid flow.

We have considered only MHVs in this paper.
The coarsening dynamics becomes different and even faster in the cases
of PCVs and one-component BECs than in the case of MHVs.
Since some of the assumptions made in this paper are invalid for PCVs
and one-component BECs, the growth laws in those cases should be
different from that of the $XY$ model or our revised one.
We will present the study about those cases somewhere else.

In conclusion, the coarsening dynamics in ferromagnetic BECs with a
positive quadratic Zeeman energy, in which magnetic anisotropy is
similar to the $XY$ model, leads to a different domain growth law from
that of the $XY$ model. 
We have proposed a revised growth law especially for the
case where only MHVs exist.
When several groups
of vortex-antivortex pairs coexist in the same space,
the difference in vortex densities of them leads to the revised growth
law. 
In the absence of the superfluid flow, where there is only one 
combination
of vortex-antivortex pairs, the growth law is the same as that of the
$XY$ model, and the coarsening dynamics is slower than in the presence of
the flow.
The effect of the superfluid flow is not only accelerating 
domain growth but also producing several combinations
of vortex-antivortex pairs.  

\begin{acknowledgments}
This work was supported by MEXT KAKENHI (No. 26103514, ``Fluctuation \&
 Structure'') and JSPS KAKENHI (No. 22740265)of Japan. YK acknowledges the
 financial support from Inoue Foundation. 
\end{acknowledgments}

\end{document}